\documentclass[aps,pra,floatfix,superscriptaddress,preprint, showpacs]{revtex4}

\usepackage{graphics}
\begin{document}
\title{Coherent control of single photon states.\footnote{Lecture given at the workshop on  "Theoretical and Experimental Foundations of Recent Quantum Technologies"
Durban, July 2006. }.}

\author{G. J. Milburn}
\affiliation{Centre for Quantum Computer technology, School of Physical Sciences, The University
of Queensland, St Lucia, QLD 4072, Australia}
%
%
\begin{abstract}
We define a class of multi-mode single photon states suitable for quantum information applications. We show how standard amplitude modulation techniques may be used to control the pulse shape of single photon states.  
\end{abstract}
\maketitle
\section{Introduction}
\label{intro}
Single photon states of light have recently received attention due to their possible application in quantum communication and computational tasks\cite{Gisin,KLM}. In this paper we wish to consider the temporal mode structure of these states and also how one might be able to control this mode structure using amplitude modulation techniques. At first sight it may seem surprising that amplitude modulation techniques would work for states that have zero field amplitude. 

The quantum electromagnetic field is described by an electric field operator\cite{Walls_Milb}, 
\begin{equation}
\vec{E}(\vec{x},t)=i\sum_{n,\nu}\sqrt{\frac{\hbar\omega_n}{2\epsilon_o V}}\vec{e}_{l,\nu}\left [e^{i(\vec{k}_n.\vec{x}-\omega_nt)}a_{n,\nu}-e^{-i(\vec{k}_n.\vec{x}-\omega_nt)}a^\dagger_{n,\nu}\right]
\label{electric-plane_wave}
\end{equation}
where $\vec{e}_{n,\nu}$ are two orthonormal polarisation vectors ($\nu=1,2$) which satisfy $\vec{k}_n.\vec{e}_{n,\nu}=0$, as required for a transverse field, and the frequency is given by the dispersion relation $\omega_n=c|\vec{k}_n|$.   The positive and negative frequency amplitude operators  are respectively $a_{n,\nu}$ and $a_{m,\nu}^\dagger$,  with bosonic commutation relations
\begin{equation} 
[a_{n,\nu}, a_{n',\nu'}]=\delta_{\nu\nu'}\delta_{nn'}
\end{equation}
with all other commutations relations zero. 

Typically we are interested in sources defined by optical cavity modes so that emission is directional and defined by the cavity spatio-temporal mode structure. Much of the difficulty in building single photon sources is in designing the optical cavity to ensure emission into a preferred set of modes. We will assume that the only modes that are excited have the same plane polarisation and are all propagating in the same direction, which we take to be the $x$-direction. The positive frequency components of the quantum electric field for these modes are then 
\begin{equation}
E^{(+)}(x, t)=i\sum_{n=0}^\infty \left (\frac{\hbar\omega_n}{2\epsilon_0 V}\right )^{1/2} a_n e^{-i\omega_n( t-x/c)}
\end{equation}
In ignoring all the other modes, we are implicitly assuming that all our measurements do not respond to the vacuum state, an assumption which is justified by the theory of photon-electron detectors\cite{Glauber}. Let us further assume that all excited modes of this form have frequencies centered on {\em carrier frequency} of $\Omega >>1$. Then we can approximate the positive frequency components by
\begin{equation}
E^{(+)}(x, t)=i\left (\frac{\hbar\Omega_n}{2\epsilon_0 Ac}\right )^{1/2}\sqrt{\frac{c}{L}}\sum_{n=0}^\infty  a_n e^{-i\omega_n( t-x/c)}
\end{equation}
where $A$ is a characteristic transverse area. This operator has dimensions of electric field. In order to simplify the dimensions we now define a field operator that has dimensions of ${\rm s}^{-1/2}$. Taking the continuum limit we thus define the positive frequency operator
\begin{equation}
a(x,t)=e^{-i\Omega(t-x/c)}\frac{1}{\sqrt{2\pi}}\int_{-\infty}^\infty d\omega' a(\omega') e^{-i\omega' (t-x/c)}
\label{field-op}
\end{equation}
where we have made a change of variable $\omega\mapsto \Omega+\omega'$ and used the fact that $\Omega>>1$ to set the lower limit of integration to minus infinity, and 
\begin{equation}
[a(\omega_1),a^\dagger(\omega_2)]=\delta(\omega_1-\omega_2)
\end{equation}
In this form the moment $n(x,t)=\langle a^\dagger(x,t)a(x,t)\rangle$ has units of ${\rm s}^{-1}$. This moment determines the 
probability per unit time (the count rate) to count a photon at space-time point $(x,t)$\cite{Glauber}. The field operators $a(t)$ and $a^\dagger(t)$ can be taken to describe the field emitted from the end of an optical cavity, which selects the directionality. 

We will contrast single photon states with multimode coherent states defined by a multimode displacement operator acting on the vacuum $D|0\rangle$,  defined implicitly by
\begin{equation}
D^\dagger a(\omega) D=a(\omega)+\alpha(\omega)
\end{equation}
where consistent with proceeding assumptions, $\alpha(\omega)$ is peaked at $\omega=0$ which corresponds to a carrier frequency $\Omega>>1$.
The average field amplitude for this state is
\begin{equation}
\langle a(x,t)\rangle=e^{-i\Omega(t-x/c)}\frac{1}{\sqrt{2\pi}}\int_{-\infty}^\infty \alpha(\omega) e^{-i\omega(t-x/c)}\equiv \alpha(x,t)e^{-i\Omega(t-x/c)}
\label{cs}
\end{equation}
which implicitly defines the average complex amplitude of the field as the Fourier transform of the frequency dependent displacements $\alpha(\omega)$
We can also calculate the probability per unit time to detect a photon in this state at space-time point $(x,t)$. This is given by
$n(x,t)=|\alpha(x,t)|^2$. Note that in this case the second order moment $\langle a^\dagger(x,t)a(x,t)\rangle$  factories, a result characteristic of fields with first order coherence. A coherent state is is closest to our intuitive idea of a classical electromagnetic field.

The multimode single  photon state is defined by
\begin{equation}
|1\rangle=\int_{-\infty}^\infty \nu(\omega)a^\dagger(\omega)|0\rangle
\label{single-photon}
\end{equation}
Normalisation requires that
\begin{equation}
\int_{-\infty}^\infty d\omega |\nu(\omega)|^2=1
\end{equation}
 This last condition implies that the total number of photons, integrated over all modes, is unity,
\begin{equation}
\int_{-\infty}^\infty d\omega \langle a^\dagger(\omega)a(\omega)\rangle=1
\end{equation}
This state has zero average field amplitude but 
\begin{equation}
n(x,t)=|\nu(t-x/c)|^2
\label{sps-count}
\end{equation}
 where $\nu(t)$ is the Fourier transform of $\nu(\omega)$. So while the state has zero average field amplitude there is apparently some sense in which the coherence implicit in the superposition of Eq.(\ref{single-photon}) is manifest. In fact comparing this to the case of a coherent state, Eq(\ref{cs}), we see that the expression for $n(t)$ is also determined by the Fourier transform of a coherent amplitude.  For this state the function $\nu(\phi)$ is periodic in the phase $\phi=t-x/c$ and it is not difficult to choose a form with a well defined pulse sequence. However care should be exercised in interpreting these pulses. They do not represent a sequence of pulses each with one photon rather they represent a single photon coherently superposed over all pulses. Once a photon is counted in a particular pulse, the field is returned to the vacuum state. A single photon state is a highly non-classical state with applications to quantum information processing. A review of current efforts to produce such a state may be found in \cite{SPS-issue}. 

\section{Manipulating single photon states.}
The relation between count rate in Eq.(\ref{sps-count}) and the Fouirer transform of the probability amplitudes in Eq.(\ref{single-photon}) indicates that if we are to engineer the pulse shape of a single photon state we must engineer the quantum probability amplitudes. While this may seem different to what we do for classical coherent pulses, it is in fact the same. The coherent amplitude displacements are also quantum probability amplitudes in a number state expansion of coherent states\cite{Walls_Milb}.  Based on this realisation it seems provident to consider how well standard coherent modulation techniques will work for single photon states. 

We first consider what happens to a single photon state reflected from an empty optical cavity: a very simple pulse shaping technique. 
In the input/output theory of quantum optics\cite{Gard_Zol}, the external field modes are related to the internal mode through the boundary condition
\begin{equation}
a_0(t)=\sqrt{\gamma}a(t)-a_i(t)
\label{boundary}
\end{equation}
Note that while $a_{i,o}(t)$ are explicitly many mode fields, the internal quasi-mode $a(t)$ is represented by a single harmonic oscillator degree of freedom with frequency $\omega_c$. The dynamics of this operator in the Heisenberg picture is given by a \begin{equation}
\frac{da(t)}{dt}=-\frac{i}{\hbar}[H_S,a(t)]-\frac{\gamma}{2}a(t)+\sqrt{\gamma}a_i(t)
\end{equation}
If these equations are linear,
\begin{equation}
\frac{d\vec{a}}{dt}=A\vec{a}-\frac{\gamma}{2}\vec{a}+\sqrt{\gamma}\vec{a}_i(t)
\end{equation}
where 
\begin{equation}
\vec{a}(t)=\left (\begin{array}{c}
					a(t)\\
					a^\dagger(t)\end{array}\right )
\end{equation}
\begin{equation}
\vec{a}_{i,o}(t)=\left (\begin{array}{c}
					a_{i,o}(t)\\
					a^\dagger_{i,o}(t)\end{array}\right )
\end{equation}
We can  Fourier transform both sides and ignore initial value terms, as we are primarily interested in the stationary statistics to obtain a linear system of algebraic equations for the frequency components of the various fields. Combining this with Eq.(\ref{boundary}) we find that
\begin{equation}
\vec{a}_0(\omega)=-\left [A+\left (i\omega+\frac{\gamma}{2}\right )I\right ]\left [A+\left (i\omega-\frac{\gamma}{2}\right )I\right ] ^{-1}\vec{a}_i(\omega)
\end{equation}
In the case of an empty cavity we get in an interaction picture defined with respect to some carrier frequency $\omega_i$ for the input external field,
\begin{equation}
a_o(\omega)=\frac{\frac{\gamma}{2}+i(\omega-\delta)}{\frac{\gamma}{2}-i(\omega-\delta)}a_i(\omega)
\end{equation}
where $\delta=\omega_c-\omega_i$. 
In other words there is simply a frequency dependent phase shift of each field mode from input to output. Suppose we input a single photon state defined by
\begin{equation}
|\psi\rangle_i=\int_{-\infty}^\infty \nu(\omega)a^\dagger_i(\omega)|0\rangle
\end{equation}
After interacting with the cavity this state is transformed to 
\begin{equation}
|\psi\rangle_o=\int_{-\infty}^\infty \nu(\omega)a^\dagger_o(\omega)|0\rangle
\end{equation}
It is then easy to see that the probability per unit time to detect a single photon in the output field is 
\begin{equation}
n(t)=\left | \int_{-\infty}^\infty d\omega\left ( \frac{\frac{\gamma}{2}+i(\omega-\delta)}{\frac{\gamma}{2}-i(\omega-\delta)}\right )\nu(\omega)e^{-i\omega t}\right |^2
\end{equation}
which is a delayed and broadened pulse. 

As the next example we consider frequency modulation inside the cavity. We assume the carrier frequency is the same as the cavity resonance frequency.  The equations of motion, in the interaction picture at the cavity resonance frequency,  are
\begin{equation}
\frac{da}{dt}=-if(t)a(t)-\frac{\gamma}{2}a(t)+\sqrt{\gamma}a_i(t)
\end{equation}
In the Fourier domain this implies
\begin{equation}
a(\omega)=\frac{-i}{(\gamma/2+i\omega)}[f(\omega)*a(\omega)]+\frac{\sqrt{\gamma}}{\gamma/2+i\omega}a_i(\omega)
\end{equation}
where $f(\omega)*g(\omega)$ is the usual convolution integral. 
if we assume that the modulation is weak, we can iterate this to first order in the modulation amplitude $f(\omega)$ to obtain
\begin{equation}
a(\omega)\approx \frac{-i\sqrt{\gamma}}{\gamma/2+i\omega)^2} [f(\omega)*a_i(\omega)+\frac{\sqrt{\gamma}}{\gamma/2+i\omega}a_i(\omega).
\end{equation}
Thus the output field is
\begin{equation}
a_o(\omega)=\left (\frac{\gamma/2+i\omega}{\gamma/2-i\omega}\right )\left [\frac{-i\gamma}{\gamma^2/4+\omega^2}(f(\omega)*a_i(\omega))-a_i(\omega)\right ]
\end{equation}
If we now assume harmonic modulation at frequency $\Omega$, so that $f(t)=\epsilon\cos\Omega t$ and
\begin{equation}
a_o(\omega)=\left (\frac{\gamma/2+i\omega}{\gamma/2-i\omega}\right )\left [\frac{-i\gamma\epsilon}{\gamma^2/4+\omega^2}a_i(\omega-\Omega)-a_i(\omega)\right ]
\end{equation}
We can now calculate the output intensity spectrum
\begin{equation}
I_o(\omega)=\langle\psi_i|a_o^\dagger(\omega)a_o(\omega)|\psi_i\rangle
\end{equation}
where the input state is defined by Eq.(\ref{single-photon}).   If we assume that $\Omega$ is much larger than the frequency width of $|\nu(\omega)|^2$, but still within the cavity line width, we find that
\begin{equation}
I_o(\omega)=\left (\frac{\epsilon \gamma}{\gamma^2/4+\omega^2}\right )^2|\nu(\omega-\Omega)|^2+|\nu(\omega)|^2
\end{equation}
In other words, excitation has been scattered into the sideband at frequency $\omega_c-\Omega$,  in the original frame of the carrier frequency, with a weight arising from the cavity response.  This is what one would expect intuitively in this situation. Note however that $\langle\psi_i|a_o(\omega)|\psi_i\rangle=0$. 

\section{Single photon code states.}
The examples of the previous section indicate that while the average field amplitude is zero, much of our intuition on amplitude modulation carriers over to single photon states  in predicting what we except to see in the field intensity (i.e. $n(t)$). We now assume that we have sufficient control over either the single photon source or subsequent optical elements that arbitrary forms for $\nu(\omega)$ (or equivalently its Fourier transform, $\nu(t)$) can be engineered.  We can now define a pulse code modulation protocol at the level of single photon states. 

Suppose we have a means to temporally modulate the amplitude function $\nu(t)$ so that
\begin{equation}
 \nu(t)=  \sum_k s_k\beta_k(t)
 \end{equation}
 where $s_k$ is a binary code symbol and $\beta_k(t)$ are the code signals\cite{Schulze-Luders}. Fourier transforming this expression implies
 \begin{equation}
 \nu(\omega)=\sum_k s_k \beta_k(\omega)
 \end{equation}
 The corresponding single photon state amy the be written as 
 \begin{equation}
 |1\rangle=\sum_k s_k b_k^\dagger|0\rangle
 \end{equation}
 where
 \begin{equation}
 b_k^\dagger =\int d\omega \beta_k(\omega) a^\dagger_\omega
 \end{equation}
which are thus seen to be a linear combination of the original plane wave field modes labelled with frequency. We can now define the single photon  code states
\begin{equation}
|1\rangle_k=b_k^\dagger |0\rangle
\end{equation}
The normalisation condition for these states is 
\begin{equation}
\int d\omega \beta^*_k(\omega) \beta_l(\omega)=\delta_{k,l}
\end{equation}
The signal single photon state, $|1\rangle$, is thus a superposition of orthogonal code states with probability amplitudes given by $s_k$, the symbols.  In coherent control the pulse codes are orthonormal with respect to signal overlap integrals. Here we see that at the level of single photons states the pulse code states are orthogonal in Hilbert space.  We anticipate that these states will play a role in single photon communication schemes.

%
%
\acknowledgements
I would like to acknowledge the support of the Australian Research Council.

\end{document}